\documentclass[aps,prb,preprint,showpacs]{revtex4}

\usepackage{graphicx}
\usepackage{dcolumn}
\usepackage{bm}

\def\ftn[#1]{\rlap{\footnotemark[#1]}}
\def\qsgw{\mbox{QS$GW$}}

\def\bfa{{\bf a}}
\def\bfS{{\bf S}}
\def\bfB{{\bf B}}
\def\bfb{{\bf b}}
\def\bfr{{\bf r}}
\def\bfq{{\bf q}}
\def\bfk{{\bf k}}
\def\bfT{{\bf T}}

\def\bfe{{\bf e}}

\def\umm{{U}^0_a}
\def\H0{H^0}
\newcommand{\req}[1]{Eq.~(\ref{#1})}
\newcommand{\ocite}[1]{\cite{#1}}

\newcommand{\ispone}{\downarrow}
\newcommand{\isptwo}{\uparrow}
\newcommand{\bfiS}{{\it \Delta \bf S}}
\newcommand{\ds}{\displaystyle}
\newcommand{\bfex}{{\bf e}_x}
\newcommand{\bfey}{{\bf e}_y}
\newcommand{\bfez}{{\bf e}_z}

\newcommand{\Ta}{{\bf T}a}
\newcommand{\Tad}{{\bf T}'a'}

\begin{document}

\title{Spin wave dispersion based on the quasiparticle self-consistent 
$GW$ method: NiO, MnO and $\alpha$-MnAs}

\author{Takao Kotani and Mark van Schilfgaarde}
\affiliation{School of Materials, Arizona State University, Tempe, Arizona, 85287-6006, USA}
\date{\today}

\begin{abstract}
We present spin wave dispersions for
MnO, NiO, and $\alpha$-MnAs based on the recently-developed 
quasiparticle self-consistent $GW$ method (\qsgw),
which determines an optimum quasiparticle picture.
For MnO and NiO, \qsgw\ results are in rather good agreement with experiments,
in contrast to the LDA and LDA+$U$ descriptions.
For $\alpha$-MnAs, we find a collinear ferromagnetic ground state in
\qsgw, while this phase is unstable in the LDA.
\end{abstract}

\pacs{71.15-m,71.10-w,71.20-Eh}

\maketitle
\section{introduction}
The magnetic linear response is a fundamental property of solids.
It is given by the spin susceptibility 
when the spin-orbit coupling is neglected (as we will do in this article).
The spin susceptibility is equivalent with the spin fluctuations, as can be seen
from the fluctuation-dissipation theorem.
Low-energy spin fluctuations can control some low energy phenomena
such as magnetic phase transitions, and
contribute to resistivity through spin-flip scattering of electrons.
Antiferro(AF)-magnetic spin-fluctuations can play an important role in
high-$T_{\rm c}$ superconductors \cite{moriya90,millis90}.  It is also the
central quantity entering into the description of quantum-critical
phenomena~\cite{Lonzarich85,Kaul99}.  
We expect that reliable first-principle methods to calculate 
the spin susceptibility should give important clues
to understand these phenomena.
In this paper, we concentrate on the magnetically-ordered systems,
where the spin susceptibility should be dominated by spin waves (SW) 
at low energy. 

In spite of the recent development of such methods,
we still have large class of systems where we can hardly 
calculate the spin susceptibility, e.g, as discussed in Ref.\onlinecite{wan06}.
A typical example is MnO; Solovyev and Terakura gave an analysis for the
calculation of its SW energies \cite{solovyev98}. Then they
showed main problem is in the non-interacting one-body Hamiltonian $\H0$ 
from which we calculate the non-interacting spin susceptibility used for the calculation 
of the SW energies. $\H0$ given by the local density approximation (LDA), 
LDA+$U$, or even the optimized effective potential (OEP) \cite{kotani98} 
are not adequate. In the LDA+$U$ case, they traced the error to a misalignment of the
O($2p$) bands relative to the Mn($3d$) bands. 
It is impossible to choose the $U$ parameter to correct 
the misalignment, because the $U$ parameter can only control 
the exchange splitting within $3d$ bands.
A possibility may be adding some other parameter in addition to $U$ 
so as to correct the misalignment; however, such a procedure including more parameters 
become less universal. 
This situation is somehow similar with the case of optical response (dielectric function)
calculation for semiconductors, where $\H0$ given by LDA is with too small band gap, 
thus requires some additional correction like scissors operator.
Our case for the spin susceptibility for MnO is rather worse; 
LDA supplies too problematic $\H0$ to be corrected in a simple manner.

Another possibility is to obtain $\H0$ by some hybrid functional;
it has been shown that it can work as explained below, 
however, it could be problematic from the view of universality.
Muscat, Wander, and Harrison claimed that a functional 
B3LYP \cite{Muscat01,becke93} (containing 20\% of Fock exchange) works even for solids.
However, Franchini, Bayer, Podloucky, Paier and Kresse 
\cite{franchini05} showed 
that a little different functional, PBE0 is better
than B3LYP in order to obtain better agreement 
with experiments as for the exchange interaction.
PBE0 is a combination of 25\% of the Fock exchange 
with a generalized-gradient approximation(GGA) \cite{perdew96}.
However, such a functional could be not so universal, 
mainly because the effect of screening 
(therefore the ratio of the Fock exchange) are dependent on materials. 
In fact, de P. R. Moreira, Illas, and Martin
\cite{moreira02} reported that a hybrid functional 
containing 35\% of Fock exchange gives best results for NiO;
the ratio of the Fock exchange is rather different from the case of 
MnO by Franchini et al.
This is somehow consistent with latest careful examinations by Fuchs 
Furthmuller, Bechstedt, Shishkin and Kresse \cite{fuchs:115109} and 
Paier, Marsman, and Kresse \cite{paier:024103};
they clarify the fact that a hybrid functional should be limited, 
because the screening effects (corresponding to the ratio of the Fock exchange) 
can be material-dependent. These seem to indicate a difficulty to pick up 
an universally-applicable hybrid functional. 
This difficulty becomes more problematic
when we treat inhomogeneous systems, e.g, to treat 
the Schotkey-barrier problem, where the screening effects
are very different in metal side and in semiconductor side.

Considering these facts, it is necessary to start from 
good $\H0$ without such problems. Our recently-developed 
quasiparticle self-consistent $GW$ method (\qsgw), which
includes the above screened exchange effects in a satisfactory manner 
\cite{Faleev04,chantis06a,vans06,chantis:165126,kotani07a,bruneval:267601,shishkin:246403}. 
\qsgw\ determines a reference system of $\H0$ 
representing optimum quasiparticle (QP) picture
in the sense of Landau-Silin Fermi liquid theory.
As discussed in Ref.\onlinecite{kotani07a},
it is based on a self-consistent perturbation theory within 
all-electron full-potential $GW$ approximation ($GW$A),
but it is conceptually very different 
from the usual full self-consistent $GW$.
\qsgw\ self-consistently determines not only $\H0$, 
but also the screened interaction $W$, 
and the Green's function $G$ simultaneously.

We have shown that \qsgw\ gives QP energies, spin moments, 
dielectric functions and so on in good agreement 
with experiments for wide range of materials. 
There are systematic but a little disagreement from experiments. 
For example, as shown in Fig.1 in Ref.\onlinecite{vans06}, we see error that
calculated band gaps are systematically larger than those by experiments.
A recent development by Shishkin, Marsman and Kresse \cite{shishkin:246403}
confirmed our conjecture \cite{vans06} that the inclusion of 
electron-hole correlation effect in $W$ will correct the error. 
Their method is a simplified version of 
the full Bethe-Salpeter equation (BSE) for $W$; 
it includes only the static and spatially-local part 
of the first-order term in the BSE, based on the procedure given by Sottile, Olevano, and Reining
\cite{Francesco03}. However, since the band-gap error itself 
is small enough, such simplifications may cause no problem.
Considering this success, we believe that \qsgw\ is a basis
for future development of the electronic structure calculations.

In this paper, we treat MnO and NiO with AF ordering II (AF-II) 
\cite{terakura84}, and $\alpha$-MnAs.  
$\alpha$-MnAs is NiAs-type grown on GaAs epitaxially, thus is a 
candidate for spintronics applications \cite{tanaka94}. 
For this purpose, we have developed a procedure to calculate 
the spin susceptibility at zero temperature.
It is a general procedure for a given self-consistent 
method which determines $\H0$, even when $\H0$ contains 
non-local potentials as in the Hartree-Fock method.
We then apply it to LDA, and to \qsgw.
After we explain the method in the next section, we will show SW 
energies obtained with \qsgw\ are in good agreement 
with experiments for MnO and NiO. 
See Ref.\onlinecite{kotani07a} for dielectric functions for NiO and MnO.
For MnAs, our calculation shows that a collinear FM 
ground state is stable in \qsgw\, though it does not in LDA.

At the end of introduction, we give a discussion to justify
using the one-particle picture (band picture) of ``Mott insulator'' 
for MnO and NiO; it is essentially given 
by Terakura, Williams, Oguchi, and K\"ubler in 1984 \cite{terakura84}
(in the following discussion, 
``charge transfer type'' or ``Mott type'' does not matter).
Based on the one-particle picture, the existence of some spin moment 
(or exchange splitting, equivalently) at each cation site 
is very essential to make the system insulator.
This is consistent with the experimental facts that
all the established ``Mott insulator'' are accompanied 
with the AF (or some) magnetic ordering.
Thus the concept ``Mott insulator vs Band insulator'' 
often referred to is misleading, or rather confusing.
In order to keep the system insulating, any ordering of spin moment is possible
provided the system retains a sufficiently large enough exchange splitting
at each site (we need to use the non-collinear mean-field method).
In this picture, metal-insulator transition at zero-temperature 
(e.g, consider a case to compress NiO) is nothing 
but the first-order transition from magnetic-phase 
to the non-magnetic phase described by a band picture.
On the other hand, the transition at finite temperature to para-magnetic phase occurs 
because of the entropy effects due to the accumulation of SWs;
then the transition is not accompanied with the metal-insulator transition
because the exchange splitting (or local moment) at each site 
is kept even above the N{\'e}el temperature $T_{\rm N}$.
This picture is very different from that
assumed in Refs.\onlinecite{kunes:156404,wan06}, where they emphasize
the priority of their method LDA + $U$ + ``dynamical mean field theory (DMFT)''.
On the contrary to their claim, we insists that our treatment 
should be prior and much closer to reality for such systems,
because of the following reasons.
\begin{quote}
i) One-particle treatment in our \qsgw\
allows us to perform parameter-free accurate calculations where we 
treat all the electrons on the same footing;
this is very critical because of the relative position of cation $3d$ bands 
to O($2p$) is important
(also their hybridization; we have no SW dispersion without hybridization).
Further, we are free from uncontrollable double-counting problem~\cite{Petukhov03}, 
nor the parameter like $U$ which is externally introduced by hand.
In contrast, LDA+$U$+DMFT carries these same problems which are in LDA+$U$,
or rather highly tangled. Thus it is better to take a calculation 
by LDA+$U$+DMFT as a model in cases.
As an example, we guess that the distribution probability of
the number of $5f$ electrons in $\delta$-Plutonium
calculated by LDA+$U$+DMFT \cite{shim07} will be easily changed 
if we shift the relative position (and hybridization) of $5f$ 
band with respect to other bands.

ii) The DMFT at zero temperature takes into account 
the quantum-mechanical onsite fluctuation which is
not included within the one-particle picture; 
it allows a system to be an insulator without magnetic order. 
However, we expect that such quantum-mechanical fluctuation 
is not essentially important to determine its ground state
for materials like NiO and MnO.
This is based on our findings that \qsgw\ results 
can well reproduce the optical response \cite{kotani07a,Faleev04},
and also the magnetic responses as shown in this paper.
These \qsgw\ results are not perfect, however,
supplies us a good enough starting point.
For example, in order to describe the d-d multiplet intra transitions
(e.g, see Fig.6 of Ref.\onlinecite{fujimori84} by Fujimori and Minami; they
are very weak in comparison with interband transitions),
it may be easier to start from the cluster models or so; however, parameters 
used in these models will be determined by \qsgw even in such a case.

iii) At finite temperature, the DMFT can take 
into account not only such quantum-mechanical fluctuations, 
but also the onsite thermal fluctuations simultaneously; 
this is an advantage of DMFT. However, in
MnO and NiO, low-energy primary fluctuations
are limited to the transverse spin fluctuations except phonons.
These can be included in DMFT but it is essentially described
by the local-moment-disorder \cite{akai93} as
the thermal average of the one-particle picture.
Thus no advantage of DMFT if only the thermal fluctuations are important.
\end{quote}

\section{method for spin susceptibility calculation}
We may divide first-principle methods to calculate SW energies
into three classes; (A), (B), and (C). 
(A) is from the Heisenberg Hamiltonian, 
whose exchange parameters $J$ are determined 
from the total energy differences of a set of different
spin configurations~\cite{Connolly83,rungger06,sandratskii06}.
(B) and (C) are based on perturbation.
(B) estimates $J$ 
from static infinitesimal spin rotations~\cite{oguchi83,Licht87}.
We go through the Heisenberg model even in (B).
In contrast, (C) determines SW energies directly from the poles in
the transverse spin susceptibility $\chi^{+-}(\bfr,\bfr',t-t')$ 
(defined below) in the random 
phase approximation (RPA) or time-dependent LDA
(TDLDA)~\cite{cooke80,savrasov98,karlsson00}.  (C) gives
the spectrum including life time, and spin-flip excitations.
Because (C) is technically difficult, (A) or (B) have been mainly used.
(B) is regarded as a simplification of (C);
but real implementations entail further approximations.

Our method belongs to (C). Our formalism is
applicable to any $\H0$ even if it contains non-local potential. 
At the beginning, we introduce some notations to 
treat the time-ordered transverse spin susceptibility 
\begin{eqnarray}
\chi^{+-}(\bfr,\bfr',t-t')
=-i\langle T( \hat{S}^+(\bfr,t) \hat{S}^-(\bfr',t') )\rangle.
\label{chipm}
\end{eqnarray}
$\langle... \rangle$ denotes the expectation value
for the ground state; $T(...)$ means time-ordering, and 
$\hat{S}^{\pm}(\bfr,t) = \hat{S}^x(\bfr,t) \pm i \hat{S}^y(\bfr,t)$
are the Heisenberg operators of spin density.
Since we assume collinear magnetic ordering
for the ground state, we have $\langle \hat{S}^x(\bfr,t) \rangle=
\langle \hat{S}^y(\bfr,t) \rangle=0$;
$2\langle \hat{S}^{z}(\bfr,t) \rangle= M(\bfr) = n^\uparrow(\bfr)-n^\downarrow(\bfr)$.
$n^\uparrow(\bfr)$ and $n^\downarrow(\bfr)$ mean up and down electron densities.
$M_a(\bfr)$ is the component of $M(\bfr)$ on the 
the magnetic sites $a$ in unit cell.
The Fourier transform of $\chi^{+-}$ is
\begin{eqnarray}
\chi^{+-} (\bfT\!+\!\bfr,\bfr',\omega)
=\frac{1}{N} \sum_{\bfq} e^{i \bfq \bfT} \chi^{+-}_\bfq (\bfr,\bfr',\omega),
\end{eqnarray}
where $\bfT$ is a lattice translation vector, and
$N$ the number of sites. $\bfr,\bfr'$ are limited to a unit cell.

Next we derive two conditions \req{summ} and \req{summ2} below,
which $\chi^{+-}$ rigorously satisfies.
Taking the time derivative of \req{chipm}, we obtain
\begin{eqnarray}
&&\frac{\partial }{\partial t'} \int d^3r'  \chi^{+-} (\bfr',\bfr,t'-t) \nonumber \\
&&= \int d^3r' \langle T([[\hat{H}, \hat{S}^+(\bfr',t')], \hat{S}^-(\bfr,t)] )\rangle
-i\int d^3r' \langle[\hat{S}^+(\bfr',t'), \hat{S}^-(\bfr,t)]\rangle \delta(t'-t),
\label{eqmotion}
\end{eqnarray}
where $[A, B]=AB-BA$. $\hat{H}$ denotes the total Hamiltonian of the system.
We have used $\frac{\partial \hat{S}^+(\bfr',t')}{\partial t'}
= i[\hat{H}, \hat{S}^+(\bfr',t')]$.
We assume $\hat{H}$ has rotational symmetry in spin space, so that
$[\hat{H}, \int d^3r' \hat{S}^+(\bfr',t')]=0$. Then the first term in the right-hand side is zero.
The second term reduces to $M(\bfr)$ because
$[\hat{S}^+(\bfr',t), \hat{S}^-(\bfr,t)] = 2\hat{S}^z(\bfr,t) \delta(\bfr-\bfr')$.
Thus \req{eqmotion} is reduced to be 
\begin{eqnarray}
\int_\Omega d^3r' \chi^{+-}_{\bfq=0}(\bfr',\bfr,\omega) =  \frac{M(\bfr)}{\omega},
\label{summ}
\end{eqnarray}
where $\Omega$ denotes the unit-cell volume. Note that \req{summ}
is satisfied for any $\omega$. 
At $\omega \to 0$, this means that $M(\bfr)$ is the eigenfunction
of $\chi^{+-}_{\bfq=0}(\bfr,\bfr',\omega)$ with divergent eigenvalue;
this is because a magnetic ground state is degenerate 
for homogeneous spin rotation.
Another condition is the asymptotic behavior as $\omega \to \infty$. It is given as
\begin{eqnarray}
\chi^{+-}(\bfr',\bfr,\omega) \to
\frac{M(\bfr)}{\omega} \delta(\bfr-\bfr') + O(1/\omega^2).
\label{summ2}
\end{eqnarray}
This can be easily derived from the spectrum representation of $\chi^{+-}$.
We use \req{summ} and \req{summ2} 
to determine the effective interaction $\bar{U}$ in the following.

As in Ref.~\onlinecite{antropov03}, 
we define the effective interaction ${U}(\bfr,\bfr',\omega)$
as the difference between $(\chi^{+-})^{-1}$ and
the non-interacting counterpart:
$\left(\chi^{0+-}\right)^{-1}(\bfr,\bfr',\omega)$;
\begin{eqnarray}
(\chi^{+-})^{-1} = \left(\chi^{0+-}\right)^{-1} + {U}.
\label{chirpa0}
\end{eqnarray}
In TDLDA, $U$ is the second derivative of the exchange-correlation energy,
$U(\bfr,\bfr')=-\delta^2E_{\rm xc}/\delta S^+(\bfr)\delta S^-(\bfr')
=I_{\rm xc}(\bfr)\delta(\bfr-\bfr')$,
which is local $U(\bfr,\bfr') \propto \delta(\bfr-\bfr')$, 
$\omega$-independent, and positive. 
Then we can show that $\chi^{+-}$ in TDLDA satisfies 
conditions \req{summ} and \req{summ2} automatically \cite{katsnelson04}. 
In the case of $\H0$ containing nonlocal potentials 
(e.g. in the case of the Hartree-Fock method), 
$U$ 
is no longer independent of $\omega$.
This is because the natural expansion of $\chi^{+-}$ 
in the many-body perturbation theory requires solving the Bethe-Salpeter Eq.
for the two-body propagator $\chi^{+-}(\bfr_1,\bfr_2;\bfr_3,\bfr_4,\omega)$,
thus ${U}$ defined in \req{chirpa0} is not 
directly identified as a kinds of diagrams.
Ref.\onlinecite{karlsson00} did not pay attention to this point.
%
We can calculate $\chi^{0+-}$ in \req{chirpa0} as
\begin{eqnarray}
\chi^{0+-}_\bfq(\bfr,\bfr',\omega) 
&&=
 \sum^{\rm  occ}_{\bfk n \ispone} \sum^{\rm unocc}_{\bfk' n'\isptwo}
\frac{
\Psi_{\bfk n\ispone}^*(\bfr)      \Psi_{\bfk' n'\isptwo}(\bfr)
\Psi_{\bfk' n'\isptwo}^*(\bfr') \Psi_{\bfk n\ispone}(\bfr') 
}{\omega-(\epsilon_{\bfk' n'\isptwo}-\epsilon_{\bfk n\ispone})+i \delta} \nonumber\\
&&+ \sum^{\rm  unocc}_{\bfk n \ispone} \sum^{\rm occ}_{\bfk' n'\isptwo}
\frac{
\Psi_{\bfk n\ispone}^*(\bfr)      \Psi_{\bfk' n'\isptwo}(\bfr)
\Psi_{\bfk' n'\isptwo}^*(\bfr') \Psi_{\bfk n\ispone}(\bfr') 
}{-\omega-(\epsilon_{\bfk n\ispone}-\epsilon_{\bfk' n'\isptwo})+i \delta},
\label{generalchi01q}
\end{eqnarray}
where $\bfk' = \bfq+ \bfk$. $\chi^{0+-}(\bfr,\bfr',\omega=0)$
is negative definite matrix.  Our definition of
$\chi^{+-}$ and also $\chi^{0+-}$ can be different in sign
from other definitions in the literature because we start from \req{chipm}.
In order to realize an efficient computational method,
we assume that the magnetization is confined to magnetic atomic sites,
and we explicitly treat only a degree of freedom of spin rotation per each site.
Then we can determine $U$ with the help of \req{summ} and \req{summ2} as in the following.
As a choice to extract the degrees of freedom, we consider a matrix $D({\bfq,\omega})$ as
\begin{eqnarray}
(D({\bfq,\omega}))_{aa'} =\int_a d^3r \int_{a'} 
d^3r'{\bar e}_a(\bfr) \chi^{+-}_\bfq (\bfr,\bfr',\omega) {\bar e}_{a'}(\bfr'),
\end{eqnarray}
and $D^{0}(\bfq,\omega)$ defined in the same manner.
The dimension of the matrix $D({\bfq,\omega})$ is the number of magnetic sites.
Here we define $e_a(\bfr) = M_a(\bfr)/M_a$ where $M_a = \int_a d^3r M_a(\bfr)$;
and define ${\bar e}_a(\bfr)$ so that ${\bar e}_a(\bfr) \propto e_a(\bfr)$
and $\int d^3r {\bar e}_a(\bfr) e_a(\bfr) =1$;
thus ${\bar e}_a(\bfr) = e_a(\bfr)/ \int_a d^3r (e_a(\bfr))^2$.
Corresponding to \req{chirpa0}, we define the effective interaction $(\bar{U}(\bfq,\omega))_{aa'}$ as
\begin{eqnarray}
(D(\bfq,\omega))^{-1} = \left(D^{0}(\bfq,\omega)\right)^{-1} + \bar{U}(\bfq,\omega).
\label{chirpa1}
\end{eqnarray}
For the calculation of $D^{0}(\bfq,\omega)$ from \req{generalchi01q},
we use the tetrahedron technique \cite{kotani07a}, which allow us to use
fewer $\bfk$ points in the first Brillouin zone (BZ)
than those required for the sampling method \cite{karlsson00}. 
$\bar{U}$ defined in \req{chirpa1} 
should include all the downfolded contributions from all the other degrees of freedom.
We now assume that $\bar{U}$ is ${\bf q}$-independent and site-diagonal, 
so that it can be written as $\bar{U}_{aa'}(\bfq,\omega)= U_a(\omega) \delta_{aa'}$.
Since \req{summ} reduces to a constraint
$\sum_{a'}  (D(\bfq=0,\omega))_{a'a} = M_{a}/\omega$,
we determine $\bar{U}_a(\omega)$ from
\begin{eqnarray}
\bar{U}_a(\omega)=\frac{\omega}{M_a} \delta_{aa'} 
-\left(\frac{\sum_{b}M_{b}(D^{0}(\bfq=0,\omega))^{-1}_{ba} }{M_a}\right)\delta_{aa'}. \nonumber 
\end{eqnarray}
With this $\bar{U}_a(\omega)$ for \req{chirpa1}, we finally have
\begin{eqnarray}
&&(D(\bfq,\omega))^{-1}=\frac{\omega}{M_a} \delta_{aa'} -\bar{J}(\bfq, \omega), 
\label{ua}  \\
&&\bar{J}(\bfq, \omega)=-\!\left(D^{0}(\bfq,\omega)\right)^{-1}
+\left(\frac{\sum_{b}M_{b}(D^{0}(\bfq=0,\omega))^{-1}_{ba} }{M_a}\right)\delta_{aa'}. 
\label{jbar} 
\end{eqnarray}
%
%
\req{summ2} reduces to
$(D(\bfq,\omega))^{-1}_{a'a} \to \frac{\omega}{M_{a}}\delta_{aa'}$ 
at $\omega \to \infty$;
$(D(\bfq,\omega))^{-1}_{a'a}$ given by \req{ua} gives this correct asymptotic behavior.
Note that we determine $U$ just from the requirement \req{summ} because
of our approximations ``onsite only $U$'' and ``a basis per magnetic site''.
If we need to go beyond such approximation (e.g. multiple basis per site),
it will be necessary to introduce additional informations, e.g. 
a part of $\chi(\bfq, \omega=0)$ evaluated by numerical linear-response calculations
(perform the \qsgw\ self-consistent calculations with bias fields).
By Fourier transformation,
we can transform $(D(\bfq,\omega))_{a'a}$ into
$D_{RR'}(\omega)$; the same is also for $D^0,J$ and so on.
Here $R=\bfT a$ is the composite index to specify an atom in the crystal.
For later discussion we define 
\begin{eqnarray}
J(\bfq,\omega)= -(D^{0}(\bfq,\omega))^{-1} +  \frac{\delta_{aa'}}{D^{0}_{aa}(\omega)},
\label{jdef} 
\end{eqnarray}
where $D^{0}_{aa}(\omega)$ is shorthand for $D^{0}_{\bfT a \bfT a}(\omega)$; 
it is $\bfT$ independent.
The second term in \req{jdef} is included just in order to remove the onsite term from $J$.
Then \req{jbar} can be written as
\begin{eqnarray}
\bar{J}(\bfq, \omega)= J(\bfq, \omega)
-\left(\frac{\sum_{b}M_{b}J_{ba}(\bfq=0,\omega) }{M_a}\right)\delta_{aa'}. 
\label{jbar2} 
\end{eqnarray}
Here, the second term (onsite term) in \req{jdef} is irrelevant because
of the cancellation between two terms in \req{jbar2}.

The preceding development for $(D(\bfq,\omega))^{-1}$ facilitates a
comparison with the Heisenberg model, whose Hamiltonian is ${\cal
H}=-\sum_{R}\sum_{R'} J^{\cal H}_{R R'}\bfS_{R}\!\cdot\!\bfS_{R'}$ ($R=\bfT
a$).  As shown in Appendix \ref{app1}, the inverse of the susceptibility in
the Heisenberg model is:
\begin{eqnarray}
(D^{\cal H}(\bfq,\omega))^{-1}
= \frac{\omega}{M_a} \delta_{aa'} - \bar{J}^{\cal H}(\bfq),
\label{dqo}
\end{eqnarray}
where $M_a=|2{\bf S}_{R}|$. 
Let us compare \req{dqo} with \req{ua}.
This $\bar{J}^{\cal H}_{aa'}(\bfq)$ is given by \req{heijbar}, 
which is almost the same as \req{jbar2}; only the difference is whether we use $J^{\cal H}$ or $J$.
This suggests how to construct the Heisenberg model which reproduces \req{ua} as good as possible; 
a possibility is that we simply assign $J(\bfq,\omega=0)$ (neglecting the $\omega$-dependence)
as $J^{\cal H}(\bfq)$. 
We have confirmed that this approximation is good enough to reproduce SW energies 
in the case for MnO and NiO.
However, it is not true in the case of $\alpha$-MnAs;
then we have used another procedure given 
by Katsnelson and Lichtenstein \cite{katsnelson04}: 
we identify $J(\bfq,\omega=(\rm SW \ energy \ at \ \bfq))$ as ${J}^{\cal H}$.
This construction exactly reproduces SW energies calculated from $D^{\bfq, \omega}$.

As a further approximation to calculate $J(\bfq,\omega=0)$,
we can expand it in real space as (omit $\omega$ for simplicity)
\begin{eqnarray}
-J_{RR'} = (D^0_{RR'})^{-1} - \frac{\delta_{RR'}}{D^0_{R}}
= (D^0_R\delta_{RR'} + D^{0,\rm off}_{RR'})^{-1}- \frac{\delta_{RR'}}{D^0_{R}}
\approx \frac{1}{D^0_R} D^{0,\rm off}_{RR'} \frac{1}{D^0_{R'}},
\label{japprox}
\end{eqnarray}
where we use \req{jdef}; we use notation that onsite part $D^0_R = D^0_{RR}$ and
the off-site part $D_{RR'}^{0 \rm off} = D^0_{RR'} -D^0_{RR'}\delta_{RR'}$.
Here we have used the assumption that $D^{\rm off}_{RR'}$ 
are small in comparison with onsite term $D^0_R$. 
This approximation corresponds to the usual second-order perturbation 
scheme of the total energy; if the spin rotation
perfectly follows the rotation of the one-particle potential,
$\frac{1}{D^0_{R}}$ is trivial; it is equal to the difference of the one-particle 
potential between spins (exchange-correlation potential in the case of DF)
because $\frac{1}{D^0_{R}}$ is the inverse linear-response 
to determine the one-particle potential for given spin rotation.
Essentially the same equation as \req{japprox} 
was used in Refs.~\onlinecite{oguchi83,Licht87}.
In cases, this approximation is somehow mixed up with the ``long wave approximation''
to expanding $J$ around $D^0(\bfq=0)$ \cite{antropov03}; however, they should be differentiated.
In order to have rough estimate of $J_{RR'}$,
we can further reduce this to the two sites model
as originally presented by Anderson and Hasegawa \cite{anderson55,mark01}.
For an AF magnetic pair (half-filled case), we obtain the following estimate:
\begin{eqnarray}
J_{RR'} \approx -\frac{1}{D_R} D^{0, \rm off}_{RR'} \frac{1}{D_{R'}}
\sim -\frac{4t^2}{{\it \Delta}E_{\rm ex} M},
\label{jest1}
\end{eqnarray}
where $t$ denotes the transfer integral, and
${\it \Delta}E_{\rm ex}$ is the onsite exchange splitting.
We have used $D_R \sim \frac{M}{{\it \Delta}E_{\rm ex}}$,
and $D_{RR'} \sim \frac{M}{{\it \Delta}E_{\rm ex}} \times
(\frac{2t}{{\it \Delta}E_{\rm ex}})^2 $.

Some additional comments.
Our formalism here is not applicable to the non-magnetic systems,
where $M(\bfr)=0$ everywhere. Then we need to determine $U$ in other ways.
A possibility is utilizing the static numerical linear-response calculations;
it gives the information of the static ($\omega=0$) part of $\chi^{+-}_\bfq$
directly (easiest spin-polarization mode at each site). 
Then it will be possible to determine $U$ from such informations
together with some additional assumptions.
In the case of systems like Gd where the $d$ shell and $f$ shell 
can polarize separately, we may need to extend our formulation
so as to include non-locality of $U$ (e.g. $U$ can be parametrized as
$U_{ijkl}$ where $i,j,k,l$ are atomic eigenfunction basis 
for $d$ or $f$ channel).

\section{result and discussion}
\subsection{MnO and NiO}
\begin{figure}[htbp]
\includegraphics[angle=0,width=0.4\textwidth,clip]{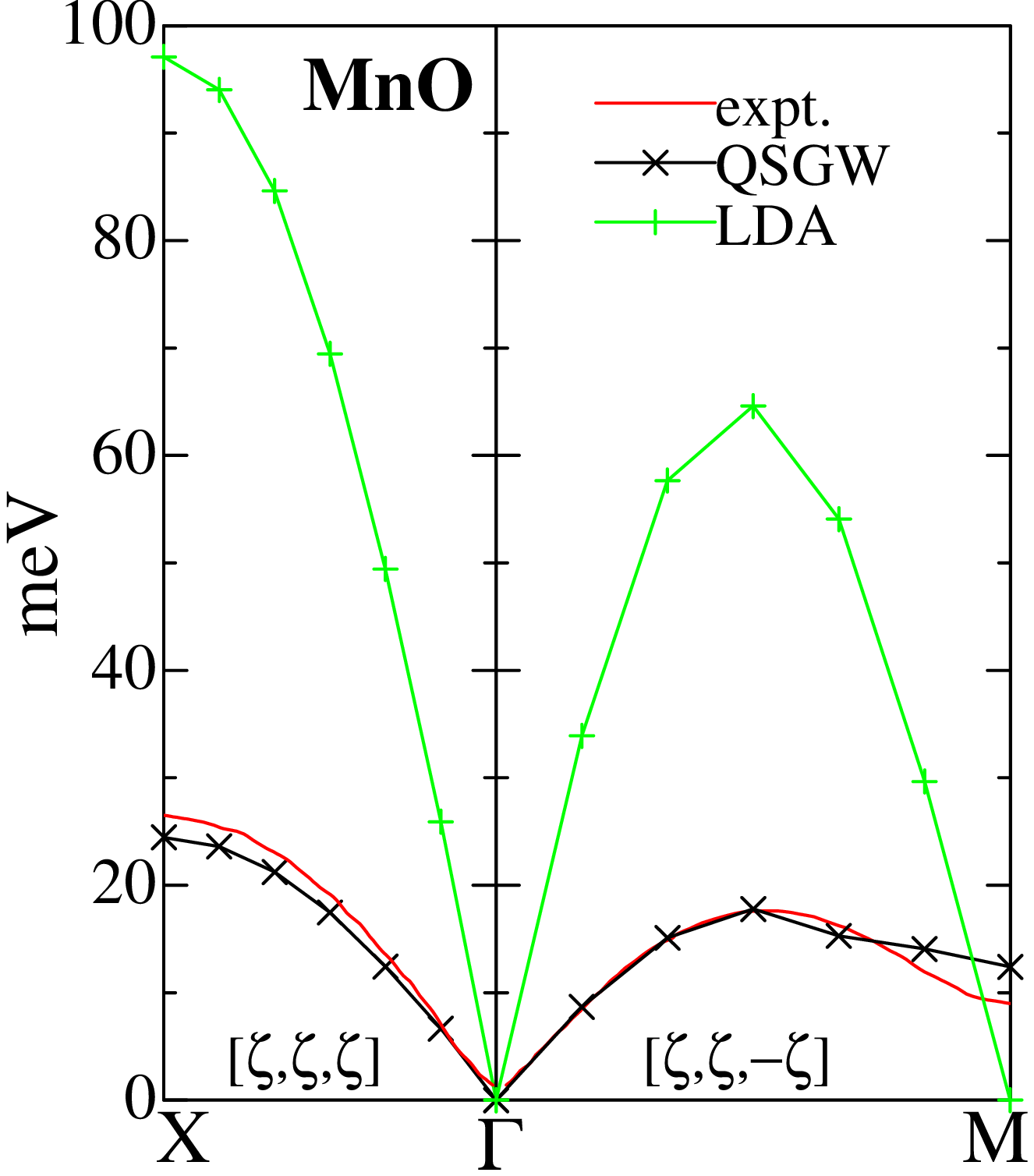}\includegraphics[angle=0,width=0.4\textwidth,clip]{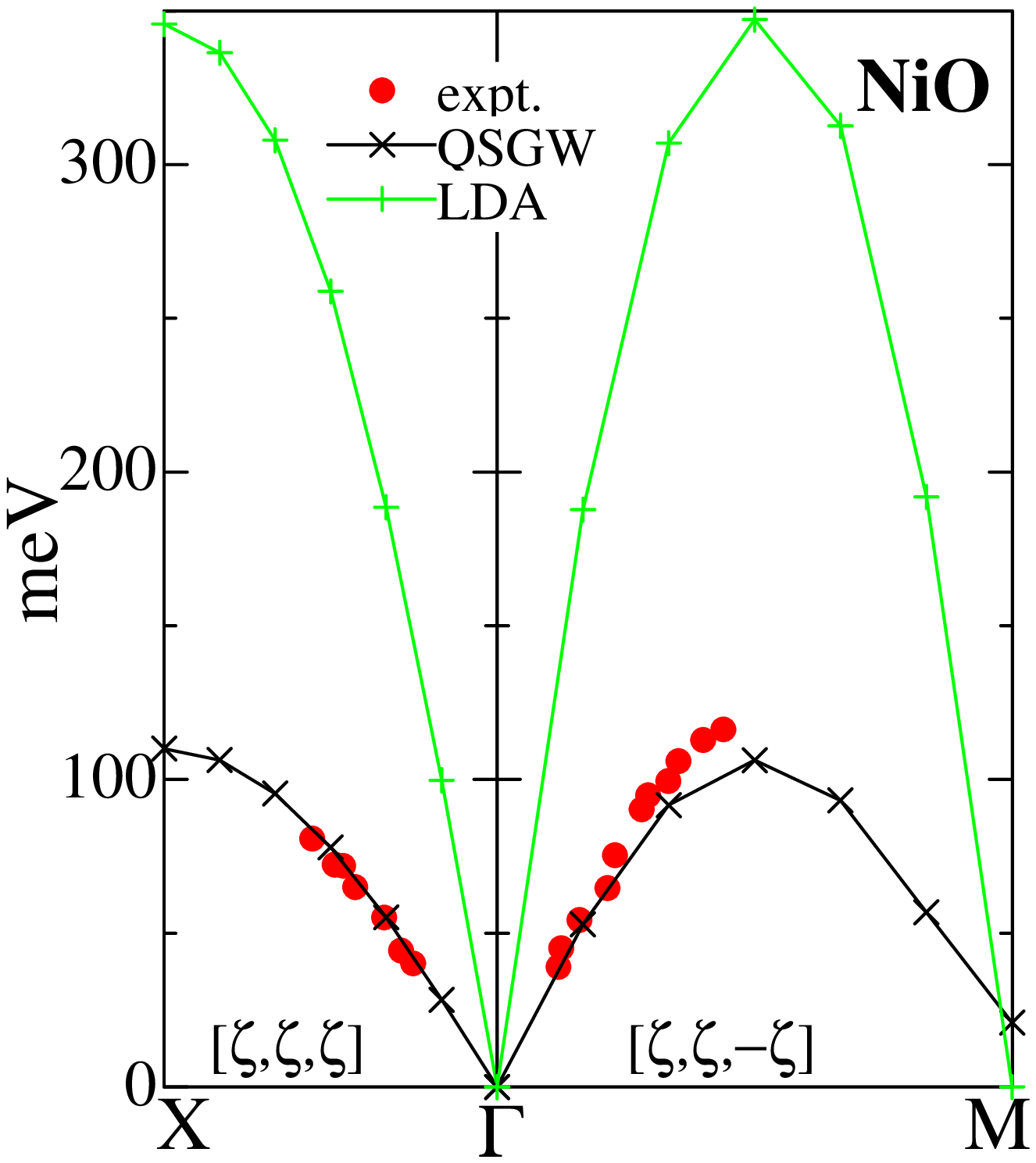}
\caption{(color online) Spin wave dispersion $\omega(\bfq)$ 
for MnO and NiO calculated from the LDA and \qsgw.
Solid line without symbols in MnO or dots in NiO (red) are experimental values
\cite{kohgi72,Hutchings72}. We used experimental lattice constants
4.55 and 4.17 \AA\, for MnO and NiO respectively.
}
\label{fig:mnonio_sw}
\end{figure}

Fig.~\ref{fig:mnonio_sw} shows the calculated SW energies $\omega(\bfq)$
for MnO and NiO.
(We used 1728 $k$-points in the BZ for all calculations, including MnAs.)
$\omega(\bfq)$ calculated from the LDA is too large, as earlier workers
have found\cite{solovyev98,wan06}.  The detailed shape of $\omega(\bfq)$ is
different from earlier work however: in Ref.~\onlinecite{wan06}, peaks in
$\omega(\bfq)$ occur near 200~meV for NiO, much lower than what we find.
\qsgw\ predicts $\omega(\bfq)$ in good agreement with experimental data.

The difference of results between \qsgw\ and LDA is understood 
by \req{jest1}. $J_{RR'}$ between nearest AF sites 
essentially determine the SW energies 
(exactly speaking, three $J$ parameters as shown in Table~\ref{tab:muu}).
The LDA severely underestimates ${\it \Delta}E_{\rm ex}$.
This can be corrected by LDA+$U$, however, Solovyev and Terakura
\cite{solovyev98} showed that it fails to reproduce SW energies 
as we mentioned in the introduction. 
This means that the transfer $t$ is also wrong in LDA+$U$; 
in fact $t$ is through the hybridization with Oxygen $2p$ 
(superexchange).
In other words, the agreements with SW experiments in \qsgw\ 
indicates that both of them are well described by \qsgw.
Together with the fact that \qsgw\ showed good agreements with optical experiments
\cite{Faleev04,kotani07a} for MnO and NiO, we claim that our one-particle 
picture given by \qsgw\ captures the essence of the physics for these systems.
Our claim here is opposite to Refs.\onlinecite{kunes:156404,wan06}
where they claimed that the one-particle picture can not capture the essence.

\subsection{$\alpha$-MnAs}
\begin{figure}[htbp]
\includegraphics[angle=0,width=0.5\textwidth,clip]{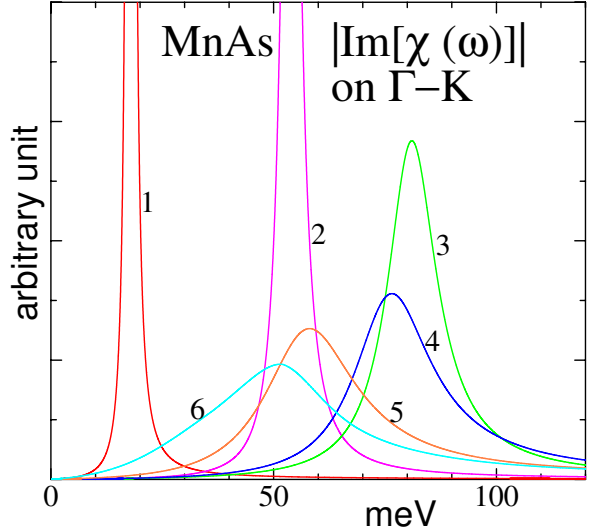}
\caption{(color online)
Im part of ${\rm Tr}[\chi^{+-}(\bfk,\omega)]$ for \qsgw.
Data are for 6 $k$-points, all along the $\Gamma-$K line.
The $k$-point is $i/6$ K (thus $i$=6 falls at K).
Peak positions and full-width-half-maxima are shown in the
$\Gamma-$K line of Fig.~\ref{fig:mnas_sw}.}
\label{fig:mnas_sp}
\end{figure}

\begin{figure}[htbp]
\includegraphics[angle=0,width=0.5\textwidth,clip]{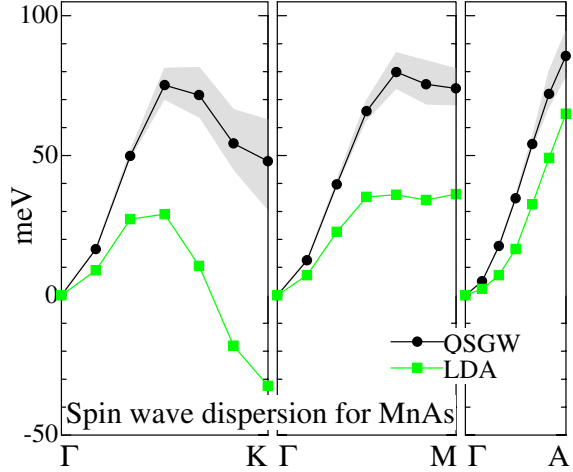}
\caption{(color online) Spin wave dispersion $\omega(\bfq)$ in $\alpha$-MnAs.  
\qsgw\ results (circles) are enveloped by hatched regions, which indicate the
full-width at half-maximum of the spin wave, and is a measure of the rate of SW decay.
LDA (squares) predicts negative SW energies around K; indicating that
the collinear FM ground state is not stable.
The experimental lattice constants $a$=3.70 \AA\ and $c/a=$1.54 were used.
}
\label{fig:mnas_sw}
\end{figure}

\begin{figure}[htbp]
\includegraphics[angle=0,width=0.5\textwidth,clip]{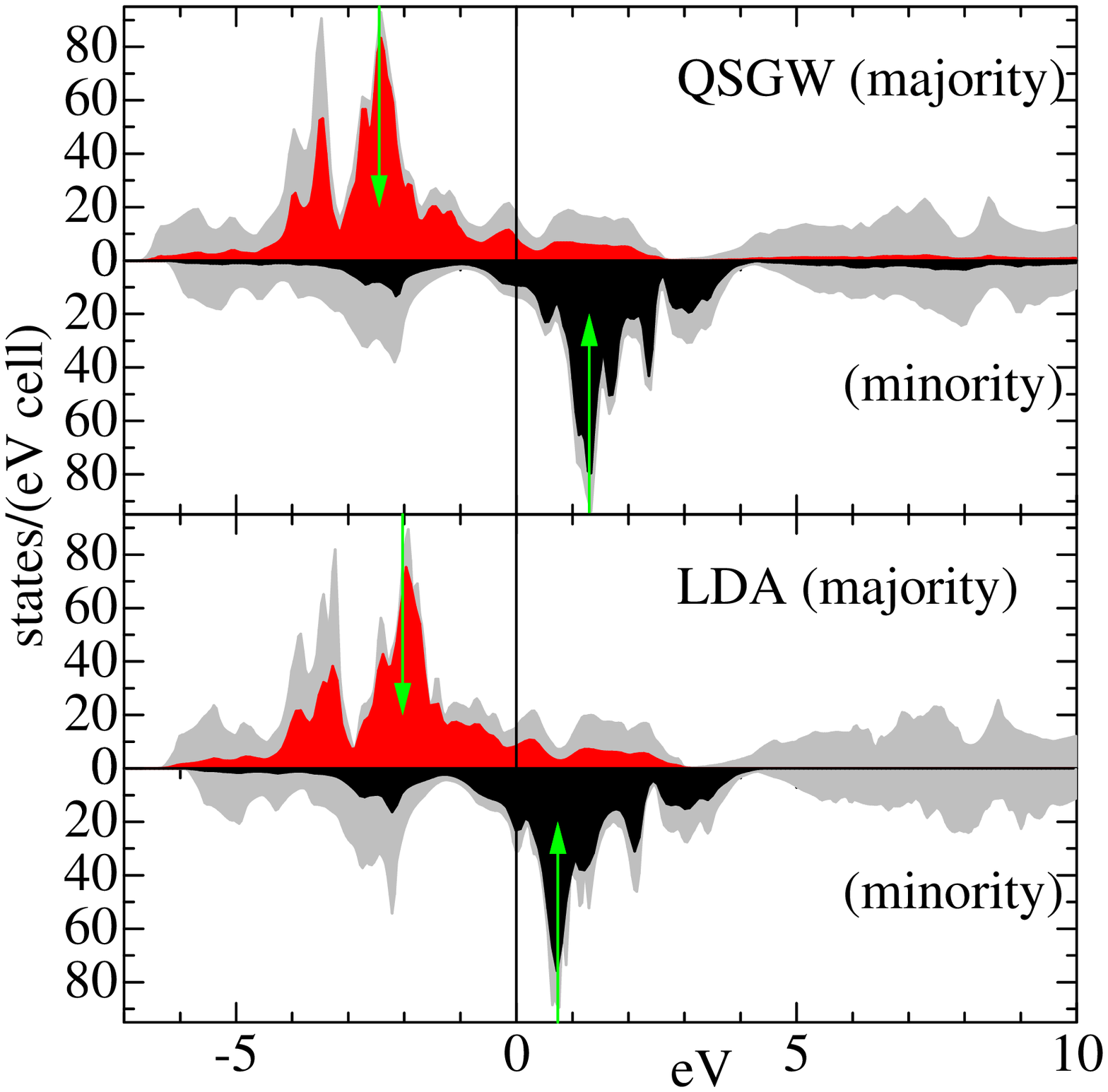}
\caption{(color online) quasiparticle DOS for $\alpha$-MnAs.
Lighter hatchings indicate total DOS; 
darker(black and red)  hatchings indicate the partial $d$ contribution, 
whose centers of gravity are shown by arrows. The Fermi energy is at zero.
}
\label{fig:mnas_dos}
\end{figure}
Because $\alpha$-MnAs is observed to be a FM with a moment of
3.4$\mu_B$\cite{goodenough67}, we construct $\H0$ assuming a FM ground state.
Inspection of the density of states (DOS) in Fig.~\ref{fig:mnas_dos}, shows that \qsgw\ predicts
${\it{}\Delta}E_{\rm{}ex}$ $\sim$~1.0\,eV larger the LDA.  This difference is
reflected in the spin moment: $M_a$= 3.51$\mu_B$ in \qsgw,
3.02$\mu_B$ in LDA.
Fig.~\ref{fig:mnas_sp} shows the imaginary part of
${\rm Tr}[\chi^{+-}(\bfq,\omega)]$ along $\Gamma-K$ line.
Sharp SW peaks are seen at small $\bfq$; they broaden with increasing
$\bfq$. Fig.~\ref{fig:mnas_sw} shows the peak positions, corresponding to
SW energies $\omega(\bfq)$.
Hatchmarks indicate the full-width at half-maximum, extracted from data
such as that depicted in Fig.~\ref{fig:mnas_sp}.
This corresponds to the inverse lifetime of a SW which
decays into spin-flip excitations.
(Our calculation gives no width for MnO and NiO, 
because of the large gap for the spin-flip excitations).
SW peaks are well identified all the way to the BZ boundary.
We find that the collinear FM ground state is not stable in the LDA: as
Fig.~\ref{fig:mnas_sw} shows, $\omega(\bfq)<0$ around K. (Among all
possible \emph{collinear} configurations, the FM state may be the most
stable.  We did not succeed in finding any collinear configuration more
stable than the FM one.  A similar conclusion was drawn for the PBE GGA
functional~\cite{rungger06}.)  On the other hand, \qsgw\ predicts stable
collinear ground state, that is, $\omega(\bfq)>0$ everywhere.  However,
even in \qsgw, the SW energies are still low around K, which is a vector
that connects nearest neighbor Mn sites in x-y plane.  If this SW energy is
further lowered for some reason, we may have a frustrated spin system
because of the triangle (honeycomb) lattice of the Mn sites.  This could be
related to the anomalous phase diagram of MnAs, which can easily occur
through the small changes in lattice structure associated with
higher-temperature phases.



We can qualitatively understand the difference 
of SW energies between \qsgw\ and LDA
from the difference of ${\it \Delta}E_{\rm ex}$.
Let us consider the energy difference 
of FM and AFM states for two-site model as illustrated in Ref.\onlinecite{mark01}.
Then the energy gain of a FM pair is independent of 
${\it \Delta}E_{\rm ex}$ when some of majority states 
are occupied (less than half filling);
we measure the energy from the majority spin's atomic level as the zero.
In contrast, the gain of a AFM pair increases with decreasing 
${\it \Delta}E_{\rm ex}$.
Overall, the LDA with its smaller ${\it \Delta}E_{\rm ex}$,
should contain a stronger AFM tendency.

\subsection{Determined Parameters and related Quantities}
\label{detpara}
%
\begin{table}[htbp]
\caption{Magnetic parameters calculated by \qsgw\ and LDA (in parenthesis).
Muffin-tin radii $R$ for cations were taken to be 2.48 (MnO),
2.33 (NiO), and 2.42 (MnAs) a.u. 
$M_a$ is the spin moment within the muffin tin.
%
%
Our approximation is equivalent to the assumption for $U$ as
${U}(\bfr,\bfr',\omega)= \sum_{a} {(U^0_{a}+\omega U^1_{a}+...) e_{a}(\bfr) e_{a}(\bfr')}$
in \req{chirpa0}. 
Then $U^0_a$ is written as $U^0_a = \int_a d^3r \int_a d^3r' e_a(\bfr) e_a(\bfr') U(\bfr,\bfr',\omega=0)$.
%
Exchange parameters $J_{1+},J_{1-},J_2$ are shown for MnO and NiO.
Total spin moments for MnAs are 7.00$\mu_B$/cell(\qsgw) and 5.89$\mu_B$/cell (LDA).
Our definition of $J_{1+},J_{1-},J_2$ follows that of 
Ref.~\onlinecite{solovyev98}, except we distinguish $J_{1+}$ and $J_{1-}$\cite{kohgi72}.
}
%
%
%

\begin{ruledtabular}
\begin{tabular}{lccccc}
              & \hfil MnO\hfil & \hfil NiO \hfil  & \hfil $\alpha$-MnAs  \\
\hline
$\umm$(eV)
             & 2.43  (0.95)  & 4.91  (1.64)  &  1.08  (0.93) \\
$M_a$ ($\mu_B$)
             & 4.61  (4.35)  & 1.71  (1.21)  &  3.51  (3.02) \\
\hline
$J_{1+}$  (meV) 
             &  -2.8 (-14.7)  & -0.77  (0.3)    &   \\
$J_{1-}$ 
             &  -4.8 (-14.7)  & -1.00  (0.3)    &    \\
$J_{2}$  
             &  -4.7 (-20.5)  & -14.7  (-28.3)  &    \\
\hline
T$_{\rm N}$ or T$_{\rm c}$ (K) &     111  & 275 & 510 &   \\
  (experiment)      & 122\ftn[1] & 523\ftn[1] & 400 
\end{tabular}
\end{ruledtabular}
\footnotetext[1]{Ref.~\ocite{roth58}}
\label{tab:muu}
\end{table}
Table~\ref{tab:muu} shows 
the effective interaction $\umm$ (interaction between unit spins).
In NiO and MnO, $\umm$ as calculated by LDA is much smaller
than the \qsgw\ result.  This is because the LDA underestimates bandgaps in
NiO and MnO, thus overestimates the screening.  
$\umm$ is twice larger in NiO than in MnO.  This
is because $M_a(\bfr)$ is more localized in NiO; in fact,
the \qsgw\ dielectric constants $\epsilon_\infty$ are similar
($\epsilon_\infty=3.8$(MnO) and $4.3$(NiO)\cite{kotani07a}),
suggesting that the screened Coulomb interaction $U(\bfr,\bfr')$
is similar in the two materials. 
$\umm$ is smaller in MnAs than in MnO, because it is a metal.

For MnO and NiO, we confirmed that $J_{RR'}$ 
is non-negligible only for the three nearest-neighbors (NN)
(Table~\ref{tab:muu}).  $J_{1+}$ and $J_{1-}$ refer to 1st NN,
spins parallel and spins antiparallel, respectively. 
$J_2$ refers to 2nd NN~\cite{kohgi72}. 
$J_{1+}$ and $J_{1-}$ by \qsgw\ are
quite different in MnO, while in LDA $J_{1+}\approx{}J_{1-}$,
resulting in $\omega({\rm M})\approx{}0$ in that case.

For MnAs in \qsgw, the expansion coefficients 
written as $(M^{-1})_{aa'}\equiv\frac{\partial (D^{\bfq \omega})^{-1}_{aa'}}{\partial \omega}|_{\omega=0}$,
is rather dependent on $\bfq$; nor is $(M^{-1})_{aa'}\propto \delta_{aa'}$.
Off-diagonal contributions of $(M^{-1})_{aa'}$ give $\sim$10 \% 
contribution to SW energies. In addition, its inverse of the 
diagonal element $1/(M^{-1})_{aa}$ is reduced by $\sim{}0.5\mu_B$ 
at certain points in the BZ. 
In this case, mapping to a Heisenberg Hamiltonian has less clear
physical meaning.

\subsection{Calculation of $T_{\rm N}$ and $T_{\rm c}$ 
based on the Heisenberg model}
From obtained $J_{RR'}$, we estimated $T_{\rm N}$ ($T_{\rm c}$
for MnAs) for \qsgw\ (Table~\ref{tab:muu}) using the cluster variation
method adapted to the Heisenberg model~\cite{Xu05}, which assumes classical
dynamics of spins under ${\cal H}$.  In NiO, the calculated $T_{\rm N}$ is
only $\sim{}50\%$ of experiment.  There are two important effects that
explain the discrepancy: $(a)$ \qsgw\ overestimates the $dd$ exchange
splitting~\cite{vans06,kotani07a}, and $(b)$ the classical treatment of
quantum dynamics of spins under ${\cal H}$.  Both effects will increase
$T_{\rm N}$.  Considering that \qsgw\ well reproduces SW energies
(Fig.~\ref{fig:mnonio_sw}), the errors connected with $(a)$ would not seem
to be so serious in MnO and NiO.  $(b)$ can be rather important, especially
when the local moment is small.  This is a general problem as discussed in
Ref.~\onlinecite{wan06}: Heisenberg parameters that reproduce SW energies
well in NiO do not yield a correspondingly good $T_{\rm N}$.  If we
multiply our classical $T_{\rm N}$ by a factor $S(S+1)/S^2 \approx 1.86$
(as $2S=M_a=1.71$), which is the ratio of quantum to classical $T_{\rm N}$
in mean field theory, we have better agreement with experiment.  This is
what Hutching et al. used~\cite{Hutchings72}.  On the other hand,
evaluation of the quantum Heisenberg model using a Green's function
technique show that the mean-field theory rather strongly overestimates
quantum corrections~\cite{GuQuantumHeisenberg05}.  Also, $T_{\rm N}$ is
already close to the experimental value in MnO. This is explained in part
because correction $(b)$ is less important in MnO, since $S$ is larger.
Further, we have large contributions to $T_{\rm N}$ from $J_{1\pm}$ in MnO,
but not in NiO.  Around $T_{\rm N}$, $J_{1+}$ and $J_{1-}$ will tend to
approach some average value, which reduces $\omega({\bf q})$ and therefore
$T_{\rm N}$ (recall $\omega({\rm M})$=$0$ when $J_{1+}$=$J_{1-}$).  The
temperature-dependence of $J$ is not accounted for here.


$J_{RR'}$ exhibits long-ranged, oscillatory behavior in MnAs: its
envelope falls off as $|{\bf R}-{\bf R}'|^3$ as predicted by RKKY
theory for a metal.  Consequently, it is not so meaningful to
estimate $T_{\rm{}c}$ from just a few NN, as was done
recently~\cite{rungger06,sandratskii06}.  Shells up to
25th-neighbors are required to converge $T_{\rm{}c}$ to within
5\% or so.  The calculated $T_{\rm{}c}$ is 110K too high in
comparison with experiment.  Taking $(a)$ into account
will improve the agreement; however, there are many factors
that make a precise calculation very difficult.
We also need to take $(b)$ into account; in addition,
other factors such as assumptions within the Heisenberg model,
may give non-negligible contributions.

In conclusion, we present a simple method to calculate spin susceptibility, 
and applied it in the \qsgw\ method.  SW energies
for MnO and NiO are in good agreement with experiments; in
$\alpha$-MnAs the FM ground state is stable, which also agrees
with experiment (to our knowledge, no SW energies have been published in
$\alpha$-MnAs).  LDA results come out very differently in each
material.  By mapping to the Heisenberg model, we estimated
$T_{\rm N}$ or $T_{\rm c}$.  We found some disagreement with
experiments, and discussed some possible explanations.

\begin{acknowledgments}
We thank M. I. Katsnelson, W.R.Lambrecht, and V.P. Antropov for valuable discussions.
This work was supported by DOE contract DE-FG02-06ER46302. 
We are also indebted to the Ira A. Fulton High Performance Computing Initiative.
\end{acknowledgments}

\appendix
\section{static $J(q)$ calculation---- Heisenberg Model}
\label{app1}
We derive the linear response to an external magnetic field $\bfB$
for the Heisenberg model, whose Hamiltonian is given as
\begin{eqnarray}
&& {\cal H} = - \sum_{\Ta} \sum_{\Tad} 
J_{\Ta \Tad} \bfS_{\Ta} \cdot \bfS_{\Tad} + g \mu_B \sum_{\Ta}
\bfS_{\Ta} \cdot \bfB_{\Ta},
\label{hei1}
\end{eqnarray}
where $\bfS_{\Ta}$ is the spin at $\Ta$ 
(${\bf T}$ is for primitive cell, $a$ specify magnetic site in a cell).
$J_{\Ta\Ta}=0$. $J_{\Ta\Tad}=J_{\Tad\Ta}$. 
The equation of motion 
$-i\hbar \dot{\bfS}_{\Ta} = [{\cal H} , {\bfS}_{\Ta}]$
is written as
\begin{eqnarray}
\hbar \dot{\bfS}_{\Ta} = \bfS_{\Ta} \times 
\left(2 \sum_{\Tad} J_{\Ta \Tad} \bfS_{\Tad} - g \mu_B \bfB_{\Ta} \right)
\label{hei2}
\end{eqnarray}
We introduce $g \mu_B \bfB = 2 \bfb$, and $\bfS_{\Ta}=\bfS_{\Ta}^0 + 
\bfiS_{\Ta}$. $\bfS_{\Ta}^0$ is the static spin configuration.
Then \req{hei2} reduces to
\begin{eqnarray}
&&\hbar \dot{\bfiS}_{\Ta} =  \bfS^0_{\Ta} \times
\left(2 \sum_{\Tad} J_{\Ta \Tad} \bfiS_{\Tad}  \right)
+ \bfiS_{\Ta} \times 
\left(2 \sum_{\Tad} J_{\Ta \Tad} \bfS_{\Tad} \right)
- 2 \bfS^0_{\Ta} \times \bfb_{\Ta} \nonumber \\
&&= \sum_{\Tad} \left(2 \bfS^0_{\Ta} J_{\Ta \Tad} \right) \times \bfiS_{\Tad}
- 
\left(2 \sum_{\Tad} J_{\Ta \Tad} \bfS^0_{\Tad} \right)  \times \bfiS_{\Ta} 
- 2 \bfS^0_{\Ta} \times \bfb_{\Ta} 
\label{hei3}
\end{eqnarray}
Introducing the Fourier transform,
$\bfiS_{\Ta} = \frac{1}{N} \sum_\bfk \bfiS_a(\bfk) e^{i \bfk (\bfT +\bfa)}$,
\req{hei3} reduces to
\begin{eqnarray}
&&\hbar \dot{\bfiS}_{a}(\bfk)   
= \sum_{a'} \left( 2 \bfS^0_{a} J_{a a'}(\bfk) 
- \left(2 \sum_{a''} J_{a a''}(0) \bfS_{a''}^0\right) \delta_{aa'}\right)  \times \bfiS_{a'}(\bfk)
- 2 \bfS^0_{a} \times \bfb_{a}(\bfk).
\label{hei4}
\end{eqnarray}
Assuming $\ds {\bfiS}_{a}(\bfk) \propto e^{-i {\omega t}/{\hbar}}$,
we have
\begin{eqnarray}
 \sum_{a'} \left(\frac{i \omega \delta_{aa'}}{2} +  \bfS^0_{a} J_{a a'}(\bfk) 
- \left( \sum_{a''} J_{a a''}(0) \bfS_{a''}^0\right) \delta_{aa'}\right)  \times \bfiS_{a'}(\bfk)
= \bfS^0_{a} \times \bfb_{a}(\bfk).
\label{hei5}
\end{eqnarray}
Let us consider the collinear ground state.  Then $\bfS^0_{a}= S_a \bfe_z$
($S_a$ is the size of spin, including sign). We have
\begin{eqnarray}
\sum_{a'} \left( \frac{i \omega \delta_{aa'}}{2S_a} \right) \bfiS_{a'}(\bfk)
+\sum_{a'} \left( J_{a a'}(\bfk) 
- \left( \sum_{a''} \frac{1}{S_a} J_{a a''}(0) S_{a''} \right) \delta_{aa'}\right)  \bfez \times \bfiS_{a'}(\bfk)
=  \bfez \times \bfb_{a}(\bfk).
\label{hei6}
\end{eqnarray}
Using $\bfS = S^+ \frac{\bfex -i \bfey}{2}
         + S^- \frac{\bfex +i \bfey}{2} + S^z \bfez$,
and $\bfez \times ({\bfex \pm i \bfey}) = \mp i({\bfex \pm i \bfey})$
we have,
\begin{eqnarray}
&&\sum_{a'} \left( \frac{\omega \delta_{aa'}} {2S_a} -
 \bar{J}_{a a'}(\bfk) \right)  S^+_{a'}(\bfk)
=  b^+_{a}(\bfk). \label{chipmh}\\
&&\sum_{a'} \left( \frac{\omega \delta_{aa'}} {2S_a} + 
\bar{J}_{aa'}(\bfk)  \right)    S^-_{a'}(\bfk)
=  b^-_{a}(\bfk),
\end{eqnarray}
where
\begin{eqnarray}
\bar{J}_{aa'}(\bfk)=  J_{a a'}(\bfk) 
- \left( \sum_{a''} \frac{1}{S_a} J_{a a''}(0) S_{a''} \right) \delta_{aa'}
\label{heijbar}
\end{eqnarray}
Only the difference between $\bar{J}_{aa'}(\bfk)$ and $J_{a a'}(\bfk)$
are diagonal parts. These are determined so that
$\int d^3k {J}_{aa}(\bfk)=0$. \req{chipmh} is the same as \req{dqo}.

\bibliography{spinwmno,gw,lmto}
\end{document}